\documentclass[conference]{IEEEtran}
\IEEEoverridecommandlockouts
\usepackage{cite}
\usepackage{amsmath,amssymb,amsfonts}
\usepackage{algorithmic}
\usepackage{graphicx}
\usepackage{textcomp}
\usepackage{xcolor}
\def\BibTeX{{\rm B\kern-.05em{\sc i\kern-.025em b}\kern-.08em
    T\kern-.1667em\lower.7ex\hbox{E}\kern-.125emX}}

\usepackage{color}

\definecolor{pblue}{rgb}{0.13,0.13,1}
\definecolor{pgreen}{rgb}{0,0.5,0}
\definecolor{pred}{rgb}{0.9,0,0}
\definecolor{pgrey}{rgb}{0.46,0.45,0.48}

\usepackage{listings}

\begin{document}

\title{Blockchain Technology to Secure Bluetooth}


\author{\IEEEauthorblockN{Athanasios Kalogiratos}
\IEEEauthorblockA{\textit{Dept. of Informatics and Computer Engineering} \\
\textit{University of West Attica}\\
Egaleo, Athens 12243, GREECE\\
cse46687@uniwa.gr}
\and
\IEEEauthorblockN{Ioanna Kantzavelou}
\IEEEauthorblockA{\textit{Dept. of Informatics and Computer Engineering} \\
\textit{University of West Attica}\\
Egaleo, Athens 12243, GREECE \\
ikantz@uniwa.gr}
}

\maketitle

\begin{abstract}
Bluetooth is a communication technology used to wirelessly exchange data between devices. In the last few years there have been found a great number of security vulnerabilities, and adversaries are taking advantage of them causing harm and significant loss. Numerous system security updates have been approved and installed in order to sort out security holes and bugs, and prevent attacks that could expose personal or other valuable information. But those updates are not sufficient and appropriate and new bugs keep showing up. In Bluetooth technology, pairing is identified as the step where most bugs are found and most attacks target this particular process part of Bluetooth. A new technology that has been proved bulletproof when it comes to security and the exchange of sensitive information is Blockchain. Blockchain technology is promising to be incorporated well in a network of smart devices, and secure an Internet of Things (IoT), where Bluetooth technology is being extensively used. This work presents a vulnerability discovered in Bluetooth pairing process, and proposes a Blockchain solution approach to secure pairing and mitigate this vulnerability. The paper first introduces the Bluetooth technology and delves into how Blockchain technology can be a solution to certain security problems. Then a solution approach shows how Blockchain can be integrated and implemented to ensure the required level of security. Certain attack incidents on Bluetooth vulnerable points are examined and discussion and conclusions give the extension of the security related problems.
\end{abstract}


\section{Introduction}

Bluetooth technology has become a part in many areas in our daily lives. From smart home appliances \cite{Hall2020} to smart cars \cite{Efstathiadis2021}, smart campuses \cite{Anagnostopoulos2021} and several other services and applications that require this technology. The existence of security vulnerabilities put at high risk the security of our data and also the security and safety of our objects in cyber-physical systems. During the process of connecting two or more devices via Bluetooth, various keys and data are generated. These keys, as well as the data, must be transferred from one device to another, in order to complete specific procedures, related to security and also the pairing process. As it is observed in several cases, during the process of transferring these keys and data, there are severe weaknesses, which can be exploited by an unauthorized user and intervene and carry out a man in the middle attack.

Blockchain technology has the necessary security \cite{Clark2021} and speed to shield the vulnerable initial communication between devices. Being a small part of the initial communication will hardly affect the speed of Bluetooth technology. The security it offers outweighs any speed cost in the initial communication. In the continuation of the communication of the devices, the Blockchain is not a part, as a result, no major changes are needed in the operation of Bluetooth. Also, the process and the steps followed by the Bluetooth to establish device connection is not affected, and remains the same, thus, maintaining the uninterrupted operation of the system. In this research work,  we identify a serious weakness in the pairing process in Bluetooth technology and propose a Blockchain solution approach presenting how the blockchain can act as an intermediary and help to the secure transfer of data and keys exchanged between two devices.

The rest of the paper is organized as follows: Section \ref{BGR} introduces the necessary background information for Bluetooth technology and Blockchain technology required to describe the problem statement and the solution concept selected to address it. Section \ref{RW} presents others' related works that expose Bluetooth vulnerabilities and mitigating approaches, addressing similar research problems.  In Section \ref{BlueVul} vulnerabilities, bugs and security issues in Bluetooth are thoroughly discussed and evaluated. Section \ref{BlockSol} presents the proposed Blockchain solution approach and then, Section \ref{Impl} demonstrates the implementation elements of the proposed scheme. An example is provided in Section \ref{Case} as a case to evaluate the implemented scheme and finally, Section \ref{CFW} discusses conclusions and future work to reveal the extension and significance of the presented research work.


\section{Background}\label{BGR}

In Bluetooth technology, two or more devices can communicate through an ad-hoc network also known as a piconet. A piconet is a Bluetooth network topology that supports up to eight devices, one of which is the \textit{master} and the rest of them are \textit{slaves}. The master device synchronizes all slave device nodes, provides the clock and performs the frequency hopping sequence within the piconet. The addressing system is designed in such a way that the master device initiates the communication link and then exchanges data with one, some or all the devices of the network.

There are three types of piconet; \textit{Point to Point}, \textit{Point to multi-point}, and \textit{Interconnection}. Brief descriptions are provided in the sequel.

\begin{itemize}

\item \textit{Point-to-Point} is the simplest piconet implementation as there are only two devices in the piconet, one is the master device and the other one is the slave device. A piconet contains only one communication link, which is between the two unique devices that exist within the network. In Fig.~\ref{figa}, the master and the slave are directly connected to each other and are the only devices in the network (Point-to-Point).

\begin{figure}[htbp]
\centerline{\includegraphics[scale=0.6]{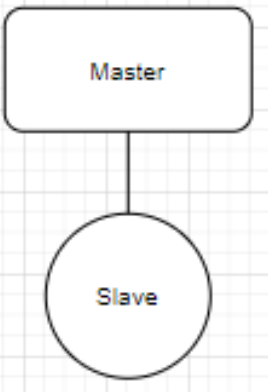}}
\caption{Point-to-Point Piconet.}
\label{figa}
\end{figure}

\item \textit{Point-to-Multipoint} contains several devices that have one device as master, while the rest are slave devices. In this case, there can be a maximum of seven active slave devices, which can only communicate with the master device and never with each other. The data rate is shared between the devices. In Fig.~\ref{figb}, a master device is directly connected to two slave devices (Point-to-Multipoint). 
\begin{figure}[htbp]
\centerline{\includegraphics[scale=0.6]{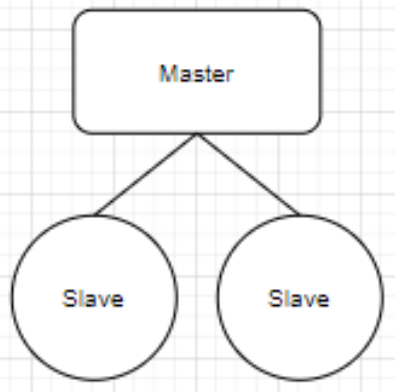}}
\caption{Point-to-Multipoint Piconet.}
\label{figb}
\end{figure}

\item \textit{Interconnection} results in the creation of a larger network called \textit{scatternet}, which supports communication between more than eight devices. In this case, a device can be a master in one piconet, a slave in another, and acts as a mediator, a bringing mechanism for exchanging data between devices that belong to different piconets. In Fig.~\ref{figc}, an Interconnection network is depicted, where master and slave devices are members of multiple networks.

\begin{figure}[htbp]
\centerline{\includegraphics[scale=0.6]{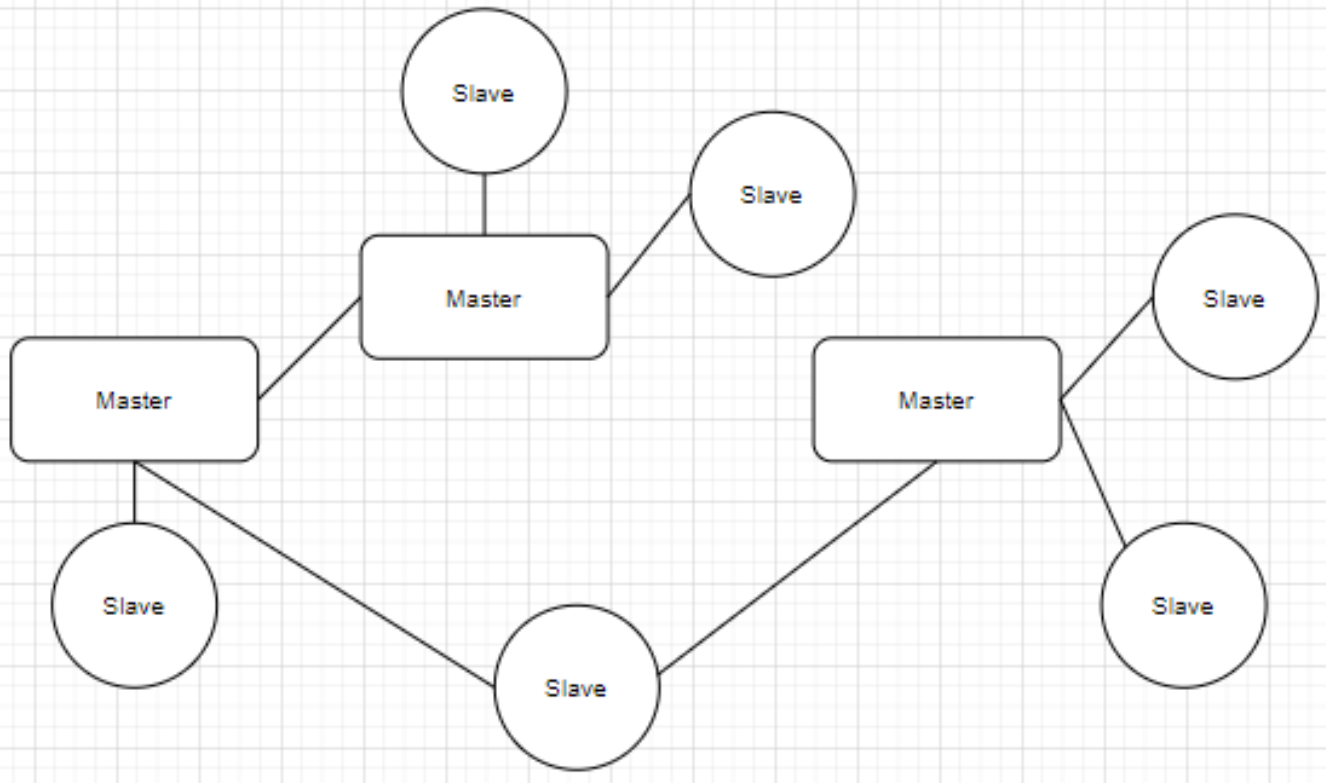}}
\caption{Interconnection Piconet.}
\label{figc}
\end{figure}

\end{itemize}



Bluetooth transmits data in packets, which are broken down into smaller packets and sent one by one to their destination. Within piconets, packets are sent directly from source to destination using the Time Division Duplexing (TDD) technique, to provide a sequential exchange of data between a master and a slave device. It can establish two different types of physical connections between devices; the \textit{asynchronous connectionless link (ACL)} and the \textit{synchronous connection-oriented link (SCO)}. The ACL is used for data exchange, while the SCO is for real-time media. In both types, data and media are transmitted in the form of packets and can be sent simultaneously, without interfering with each other during transmission.

\begin{figure}[htbp]
\centerline{\includegraphics[scale=0.55]{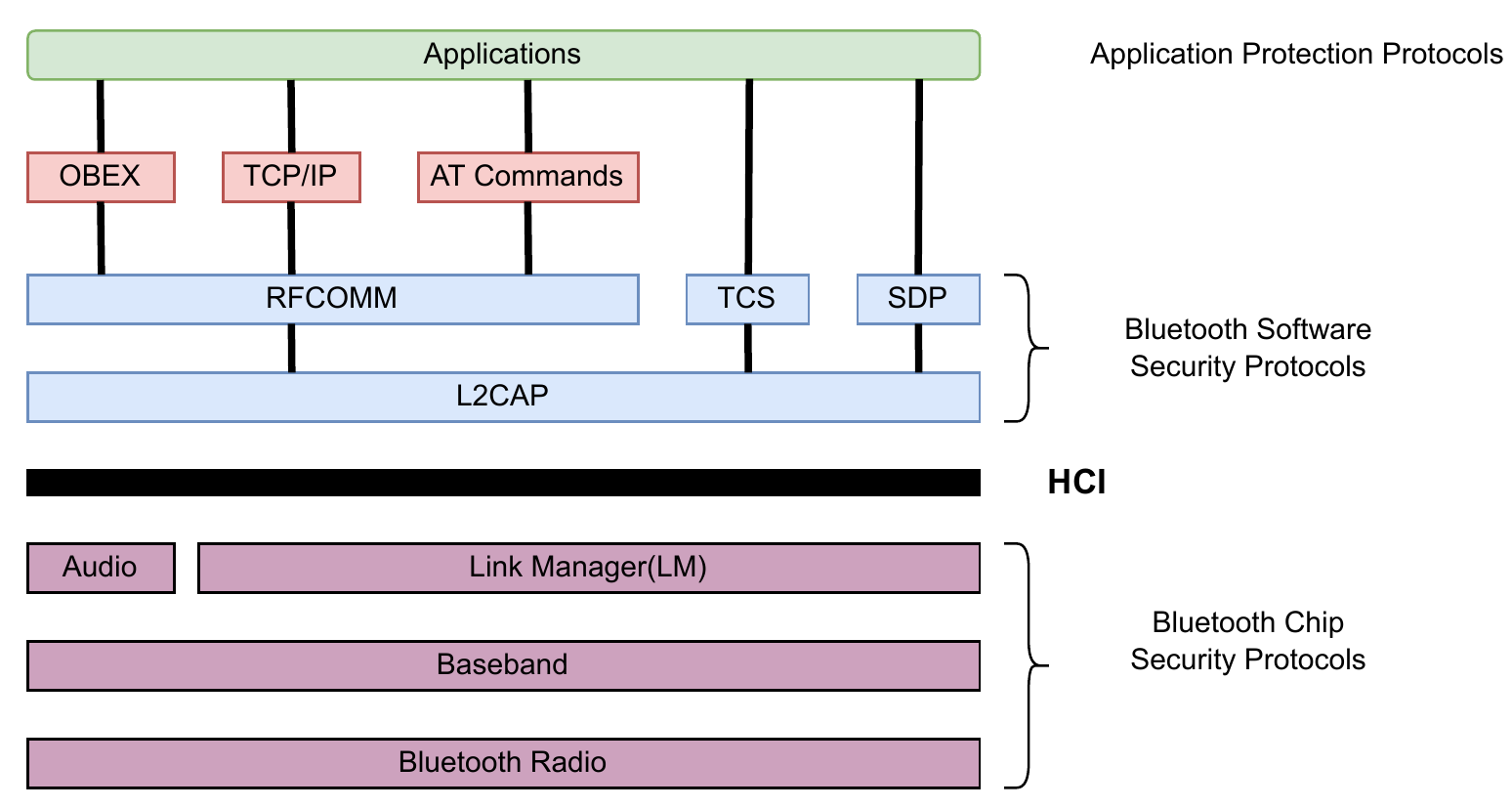}}
\caption{Bluetooth Architecture.}
\label{arch}
\end{figure}

The Bluetooth system is inside a chip and has a specific architecture, responsible for the communication of the devices. In order to be used by a device, the device must support some profiles, necessary for communication with other devices. The architecture of Bluetooth is depicted in Fig.~\ref{arch}. Complete detailed Bluetooth specifications are given in \cite{Wooley2020}.

\section{Related Work}\label{RW}

In \cite{Be-Nazir2012} an overview of Bluetooth vulnerabilities are described among with some proposed solutions. The authors refer to attacks that have been carried on the past. They describe the attacks as long as the vulnerabilities that allow the attacks to take place. A matrix that contains all the Bluetooth security vulnerabilities with their descriptions. The research concludes that the weakest part of the Bluetooth technology involves the pairing process, in which it establishes  trusted  relationships  with other devices. Then a table is provided, with possible countermeasures in order to mitigate the risk of exploiting such vulnerabilities. Since then, many new attacks have been recorded, as the range of devices using Bluetooth has massively been increased.

Reference \cite{Lonzetta2018} reviews Bluetooth security issues while discussing numerous Bluetooth threats and vulnerabilities. This article also imputes the process of pairing as not being secure. Thus, at this stage we find the Bluetooth to be vulnerable to attacks. Additionally, risk mitigation is examined with some countermeasures. Finally real-life exploitations and solutions are presented, also recommending security measures, in order to make Bluetooth communications more secure.

We can see some critical problems and risks being described in \cite{Hassan2017} that have been identified in Bluetooth. This research paper names some attacks, explains how they work and what are the consequences of them. Finally, it points out some threats that need to be sorted out in order for the Bluetooth to be more secure.

In the research papers above the authors analyze a total of sixteen different attacks such as MAC spoofing attack, PIN cracking  attack, Man-in-the-Middle/Impersonation  attack, BlueJacking  attack, BlueSnarfing attack, BlueBugging  attack, BluePrinting  attack, Blueover attack, off-line PIN recovery attack, brute-force attack, reflection attack, backdoor attack,  DoS attack, (14) Cabir worm, Skulls worm, and Lasco worm, with the most dangerous being Man-in-the-Middle(MITM)/Impersonation attacks. The conclusion of the literature review is that the pairing process and the early stage of connections in general are vulnerable. The papers also recommend solutions that cover some of the vulnerabilities. Therefore, in this paper the early stage vulnerabilities are analyzed and a workaround solution that incorporates the Blockchain technology is proposed.


\section{Bluetooth Security and Vulnerabilities}\label{BlueVul}

The core Bluetooth security mechanisms are provided and discussed in the next subsection, to reveal their strengths and weaknesses and the provided services. Existing vulnerabilities and frequent attacks are exposed subsequently, and evaluation of their significance justify the proposed solution approach.

\subsection{Fundamental Bluetooth Security Mechanisms}

The Bluetooth protection model contains four accurately executed processes that are necessary for the secure connection of devices and for file transfer.

The first process is called \textit{pairing} where the necessary secret keys are created. This pairing only takes place the first time two devices come into contact. In case it has already been done, the connection process starts with the verification process. The pairing process consists of the steps described in the following.

\begin{itemize}

    \item \textit{Initialization Key Creation}: The initialization key is always set up when any two devices get into contact for the first time. Each device comes with a unique Bluetooth address from the factory, in the size of 48bit, and we refer to it as BD\_ADDR. To generate the initialization key, a random number is sent from device A to device B, where B has approached A for data exchange. Device B generates the initialization key using a random number, the Bluetooth address, the PIN provided by the user, and a specially crafted algorithm.
    
    \item \textit{Link Key Creation and Exchange}: To create the Link key, the initialization key is used. Device A sends a random number to device B, which generates the initialization key. Device B also generates the Link key. The Link key is generated from the B's Bluetooth address and another random 128bit number, using a certain algorithm. When the Link key is created, it is entered as input in a XOR function along with the initialization key, and the output is sent to device A. Now device A can calculate the initialization key since it has the random number it generated, B's address and the pass key (PIN) and also get the Link key by performing the XOR property.
    
\end{itemize}

The bonding process follows the pairing process, during which the keys created are stored, so that, they can be used later for various types of connections in order to create a trusted pair of devices.

Immediately after the bonding process, the authentication process comes. It consists of a verifier and a claimant. Device A is the verifier and device B is the claimant. For verification purposes, device A challenges device B and waits for a response. Device A sends a random 128-bit number to device B. Device B uses its address, the random number and the Link key and calculates a response, which it sends back to device A. If it matches device A's calculated response, then the two devices are each authenticated with the other. This step is performed after the pairing process, or in the future, without performing the Pairing process again, as the necessary keys have already been created. The Link key, which has been generated and used for the pairing process, will be used as a parameter to generate the encryption key.

Finally, after the pairing, the bonding and the authentication of the devices is completed, it's time to encrypt and transfer files. The Encryption key is generated using the Link key that has been shared through the authentication process, a 96bit offset and a 128bit number. A certain Bluetooth algorithm is used to generate the Encryption key and then the Encryption key is used to encrypt the data to be transferred.

\begin{figure}[htbp]
\centerline{\includegraphics[scale=0.08]{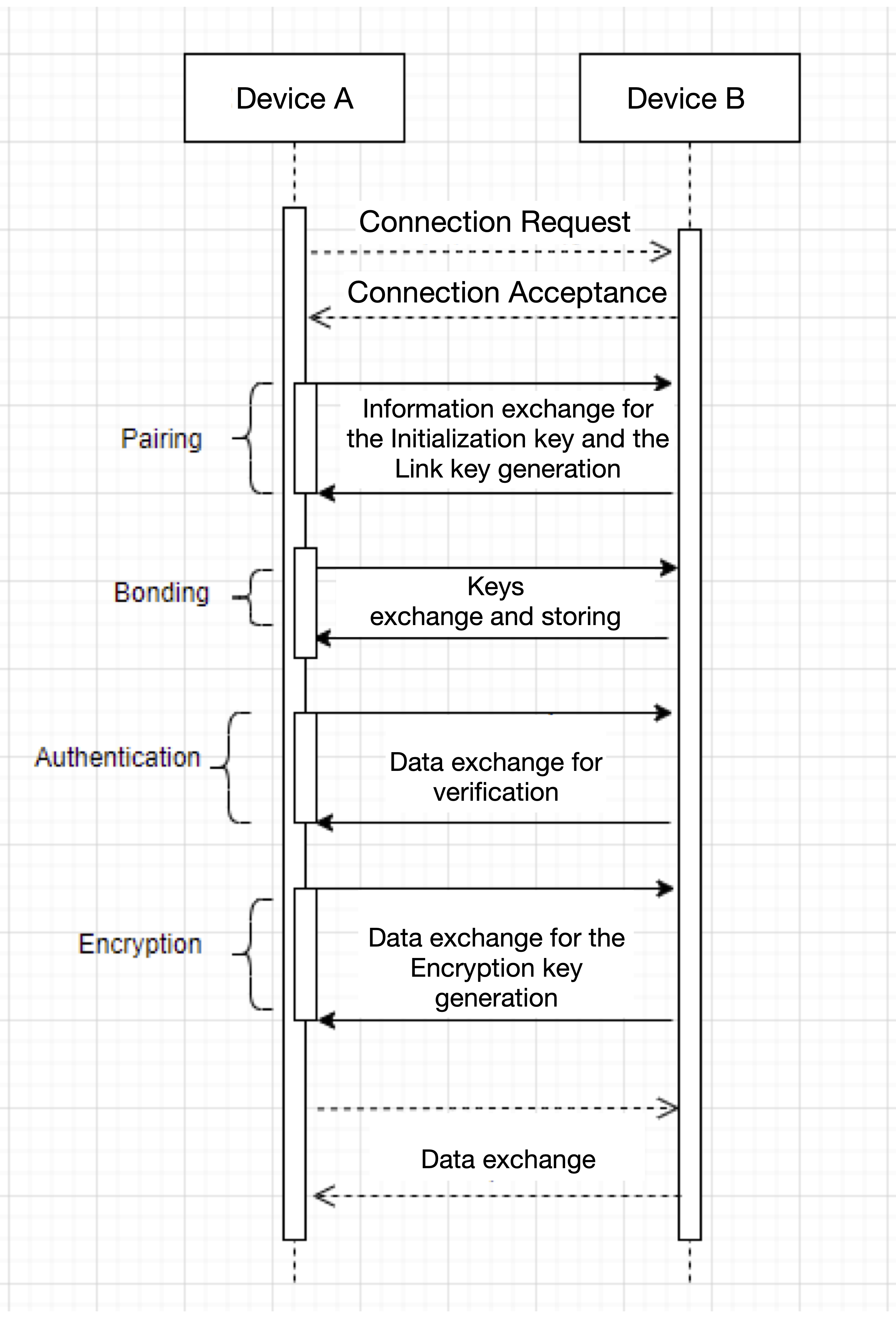}}
\caption{Pairing, Bonding, and Authentication Processes.}
\label{4stages}
\end{figure}

Figure~\ref{4stages} illustrates the walk-through of the processes coupling, bonding, verification, encryption and file transfer.

\subsection{Bluetooth Vulnerabilities and Attacks}

When connecting two devices using Bluetooth technology the following stages are performed in order: \textit{Pairing}, \textit{Bonding}, \textit{Device Authentication}, \textit{Encryption}, \textit{Data Exchange}. Pairing, Bonding, and Device Authentication, discussed before, are the fundamental security mechanisms in Bluetooth technology.

Prior to the Encryption stage, communication between devices is not protected in any way. This gap facilitates a malicious user to interfere with device communication. Due to this weak point, Bluetooth is vulnerable in these first communication stages and it is noticeable that additional vulnerabilities can be identified and most attacks mainly aim at the Pairing stage, where there is a critical exchange of security information.

This weakness during the connection has been the focal point of different incidents. In \cite{Antonioli2020b} the researchers exploit a vulnerability regarding the entropy of encryption keys during the negotiation stage. We can also see in \cite{Antonioli2020a} that a vulnerability has been spotted during the establishment of secure connection. Both of these incidents take place during the pairing/bonding stage.

Below we examine two serious vulnerabilities that should be addressed. By analyzing them we identify the stage of communication in which Bluetooth is vulnerable.

\begin{itemize}

    \item{\textit{Key Negotiation of Bluetooth Attacks}}: The Key Negotiation of Bluetooth (KNOB) attack takes advantage of the low entropy used in the negotiation of encryption keys. It is allowed for a third party to interfere the negotiation without knowing any other key and make the two device to agree on encryption keys with 1 byte of entropy and then brute force the keys in real time. By knowing the encryption keys, a third party can easily decrypt the data, which are being exchanged and either steal or alter them. This attack cannot be realized by the victims, because the whole process of key entropy negotiation is taking place in the background.  All the versions of Bluetooth are vulnerable to this attack, as they all support encryption keys with 1 to 16 bytes of entropy, and in addition, there is no security mechanism in the protocol that is being used. The target of this particular attack is the software of the Bluetooth chip, the Bluetooth controller, as all security functions are carried out there.

    \item{\textit{Bluetooth Impersonation Attacks}}: The "Bluetooth standard" contains two authentication processes, a \textit{legacy} and a \textit{secure} one. Both processes can be used on devices to identify one another using a long-term key. These procedures take place during the pairing process to prevent masquerade attacks, known as Bluetooth Impersonation Attacks (BIAS). However, there are some vulnerabilities that allow an attacker to carry out impersonation attacks, during the establishment of a secure connection. Such vulnerabilities include the lack of mandatory confirmation, the overly easy exchange of roles, and the deterioration of the confirmation process. Like KNOB attack, the BIAS attack is invisible to the user, because it does not involve him anywhere, nor does it affect his interaction with the device.
\end{itemize}

Studying these two vulnerabilities, as well as vulnerabilities that had arisen in the past, we recognize that all the sensitive points of the bluetooth system are the processes that take place during the connection between devices, and specifically, during the process of Pairing and Authentication. In these processes the system exchanges valuable and sensitive information which play a very important role in the security and integrity of the system. The security systems that Bluetooth has do not achieve to cover all the weaknesses of the system. For this reason, we introduce the Blockchain technology as an appropriate and effective solution approach to this problem (\cite{Abdelgalil2021}, \cite{Dimitriadis2021}), and we explain how the pairing process can be performed in a secure environment of a Blockchain system.

\section{Blockchain Solution Approach}\label{BlockSol}

Blockchain technology can be considered as a public ledger in which transactions are recorded and stored in blocks. In such a way, a chain is created and grows each time new blocks are added. Security has been added to this system, by the inclusion of asymmetric cryptography and distributed consensus algorithms, which also help with the coherence of the ledger. What sets blockchain technology apart from other practices is the characteristics of decentralization, consistency, anonymity, immutability and control. Since this technology allows a transaction to be completed without any intermediary, it is suitable for financial transactions too. But, it can also be used for other purposes, mainly in the form of private blockchain, such as contracts, public services, security services, as well as for the Internet of Things. Blockchain can act as an intermediary and help in the secure transfer of data and these keys exchanged in Bluetooth pairing and authentication processes.

\begin{figure}[htbp]
\centerline{\includegraphics[scale=0.5]{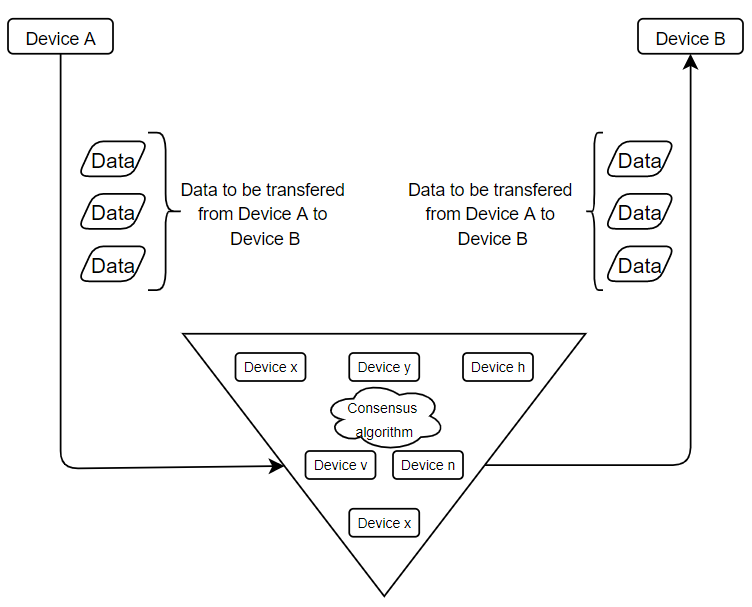}}
\caption{Blockchain Solution Approach.}
\label{fig}
\end{figure}

The diagram in Fig.~\ref{fig} illustrates the proposed solution approach with the use of a blockchain system, which is described in details in the following.

Each device has generated a pair of keys, as part of an asymmetric cryptographic algorithm, one is the public and the other one is the private. In our case, the RSA algorithm has been selected and used, but any other asymmetric cryptographic algorithm can be used, with corresponding characteristics. A private Blockchain system includes devices of an IoT \cite{Gerodimos2021}. Then all the devices that belong to this Blockchain system disclose their public keys, while keeping the private keys hidden and secure. A public key can be used, by the sending device, to encrypt the information to be transferred, so that, no other device on the blockchain network can view it. Correspondigly, a private key can be used by the recipient of the information to decrypt it. Using this cryptographic scheme, confidentiality of the data transferring is ensured, because the receiver decrypts data with the private key, which is secretly kept and only she knows.

The sender then proposes a new block to be added to the network based on the sender, the recipient, and the information he wants to transfer. This information as mentioned above is information that plays an important role in the pairing process (eg keys, degree of entropy, etc.). The block also contains the hash of the previous block, as well as its own hash. This block is then added to the blockchain system, when the consensus algorithm process is complete. The blockchain data can be accessed and viewd by any member of the blockchain system, as he keeps a copy of the blockchain locally, and this is what makes the system, immutable and so secure. However, as mentioned above, the data is encrypted and can only be decrypted by the recipient, as he holds the private key required for decryption. The recipient can then decrypt the data and process it as required and specified by the Bluetooth protocol. This process can be repeated for any device that belong to this blockchain system as member.

\section{Implementation}\label{Impl}

Below we will see the implementation of a program that has been designed and implemented to simulate the application of blockchain technology in functions that require Bluetooth communication. The principle idea is that Bluetooth uses secure blockchain technology to carry out the pairing process, which has several vulnerabilities. More specifically, critical information or data such as keys, entropy, codes, etc. are exchanged, which contribute to the secure connection of the devices via Bluetooth.


The program is written in Java language and consists of 4 classes.

\subsection{"Block" class}

The "Block" class is responsible for implementing Blockchain Blocks. Each Block contains 3 variables. The hash of the previous block (previousHash), the transmitted data (data) and the hash of the block itself (blockHash). Then a constructor is implemented, in which we pass the data  as input, the hash of the previous block. Then the hash of the block itself is calculated using the hash that was accepted as input. Finally, we have created some 'get' methods, so we can export the values of the variables previousHash, data, blockchain.

\subsection{"RSAKeyPairGenerator" class}

The "RSAKeyPairGenerator" class is responsible for generating the keys of the asymmetric cryptographic algorithm (publicKey and privateKey), with which the encryption and decryption are performing respectively. These keys are used also in the digital signature function. In the RSAKeyPairGenerator method, a type of RSA keys of size 1024 bit is used. The key generation algorithm is a class embedded in java by oracle and is located in the java.security library. This is followed by a method that is responsible for storing the keys in a file. Finally, two methods of 'get' type are also included, so that we can retrieve the private and public key.

\subsection{"RSAUtil" class}

The "RSAUtil" class contains methods for encrypting and decrypting keys (private and public) and block data. Java libraries are used for their implementation. The getPublicKey and getPrivateKey methods are responsible for encrypting the keys to make them secure. The encrypt and decrypt methods are used to encrypt and decrypt data.

\subsection{"Main" class}

The "Main" class is the principal class through which all the steps that would follow a transaction in the blockchain between two devices are performed. First, we create a dynamic size table, which will be our blockchain. Next, we observe the production of the keys used to encrypt and decrypt the data. This step in real environment will be performed separately on each device, and then only the public key will be announced. Next, the data are stored encrypted in a variable. We then enter the encrypted data into a block, which waits to be confirmed by the consensus algorithm and entered into the blockchain.

Data encryption and block request are all completed locally on the device. The corresponding device that receives the data then decrypts the data with the private key and uses it accordingly. The data are visible on all devices, but they are encrypted, so only the device whose public key was used can decrypt and view the data.

Since the blockchain system is closed and secure, the data could also not be encrypted and seen by all nodes connected to the respective device. This way, few transactions will be made without the need of computing power, each time for a new transaction. Nevertheless, encryption is an excellent extra security on the system.

\section{Case Study}\label{Case}

An example of running the program is provided and at the same time we explain at each step what is happening, why and how security enhancements are achieved. Because communication between devices using Bluetooth technology before the encryption stage is not secure, Blockchain technology is used from the beginning. Blockchain technology does not replace the device pairing process but is added on to it. The pairing process is performed normally, with the difference that the data is exchanged through the Blockchain system and not wirelessly as before.

\begin{footnotesize}
\begin{lstlisting}
//Creating a new pair of private-public keys.
RSAKeyPairGenerator KeyPairGenerator = new RSAKeyPairGenerator();

//Setting the private key in a usable variable.
String PrivateKey = Base64.getEncoder().encodeToString(KeyPairGenerator.getPrivateKey().getEncoded());

//Setting the public key in a usable variable.
String PublicKey  = Base64.getEncoder().encodeToString(KeyPairGenerator.getPublicKey().getEncoded());
\end{lstlisting}
\end{footnotesize}

The case starts from each device, in which the production of a pair of keys takes place, the private and the public. The private is kept confidential, while the public is disclosed to the other devices in any way.

While the data are still on the device, they are encrypted by the corresponding algorithm, as shown below.

\begin{footnotesize}
\begin{lstlisting}
try {  //try and catch exceptions
//Encryption
String Data = "Key1=55654415, Key2=5665415564";

//Sender encrypts data with the public key of the receiver.
String encryptedString = Base64.getEncoder().encodeToString(RSAUtil.encrypt(Data, PublicKey));
System.out.println("Encrypted data: " + encryptedString);
\end{lstlisting}
\end{footnotesize}

Then, the encrypted data enter the Block, which in turn will enter the "chain". Bluetooth would send the data in plaintext and "open" in the sense that they are not additionally protected by a network, as done in our case. At this point, the Hash of the Block is created according to a combination of the data provided by the device, and the Hash of the previous Block. In this way, all the Blocks are connected, and so it is impossible to change the Block data. This means that if the data of a Block are changed, then the Hash of all other Blocks is changed too, and so a malicious action can be detected and rejected by the system.

\begin{footnotesize}
\begin{lstlisting}
String genesisData = encryptedString;              //Data transferred.
Block genesisBlock = new Block(genesisData, 0);    //Creating new block.
System.out.println("Block hash: " + genesisBlock.getBlockHash());
\end{lstlisting}
\end{footnotesize}

Subsequently, in a real system a consensus algorithm would follow. This algorithm can be any of the algorithms mentioned above. In our case, any algorithm could be used, because it wouldn't have any special meaning.

Finally, the device that receives the information, locally decrypts it with its private key and confirms that the data come from the device which needs them, and uses them accordingly. In a real implementation, the necessary software would be integrated, so that it would be obvious to which device the data go, using the Public key of the receiver.

\begin{footnotesize}
\begin{lstlisting}
//Decryption
String decryptedString = RSAUtil.decrypt(encryptedString, PrivateKey); //Receiver decrypts data with his private key.
System.out.println("Decrypted data: " + decryptedString);
    } catch (NoSuchAlgorithmException e) {
        System.err.println(e.getMessage());
}//end of try catch
\end{lstlisting}
\end{footnotesize}

The data transaction remains in the system forever and helps in the continuous protection of the proposed system. In the frame below, a sample of some information are portrayed, as produced during the operation of the program.

\begin{footnotesize}
\begin{lstlisting}
//Encrypted data:
aMTq1IIPWAPPtpwcr4Obs5dz7StKufp84hWHwRERcyUTQZmYzl8
IVeq6wlviTWkkS0gIIRvUjiurqdXbvPJx06m4AYrw4Pm+GGgyvbR6P+7/NfttJ4fxUKG4+iXCsx4clumWRee59k1Nz9EhOUXMUFBi/54JDGolWxZW2IR1lOc=

//Block hash:
-1940134471
//Decrypted data:
Key1=55654415, Key2=5665415564
\end{lstlisting}
\end{footnotesize}

In the first and second line, the information/data in encrypted form are presented. At this stage, the keys have been created, the public and private keys have been exchanged, and the data have been encrypted. In the next lines, the hash of the block, created in the blockchain, is located. Finally, the data after this follow, decrypted with the use of the private key. All these data are printed only for our case, to facilitate the better understanding of the various stages of transferring. In a real environment, no information or data are used by the Bluetooth protocol nor by the blockchain network in a printing form.

\section{Conclusion and future work}\label{CFW}

An IoT system has a great number of beneficial elements to gain from Blockchain technology. The combination of security and speed is suitable, as a key feature in the Bluetooth security system, which is the main intermediator between IoT communicating devices. Bluetooth technology suffers from a lack of protection for the original communication schemes used for devices. At this stage, critical information is exchanged, which should be protected. This can be achieved by the use of Blockchain technology, which can shield the initial information transactions.

Further study is required towards the use of the appropriate Network Access Control system, in order to import devices into the Blockchain network in a secure way. This step is really significant, because if the system does not provide high level security, then the network is at severe risks and compromised, putting the protection provided by the Blockchain technology into question.

Certain improvements could be made to advance the implementation of the proposed system, and simulate real environments with Blockchain technology. In addition, other improvements include encryption enhancement and making encryption more efficient, both by changing the key generation algorithm and by using the appropriate consensus algorithm. In terms of Blockchain security, more block features could be embedded and used to create a more powerful block hash.

\end{document}